\documentclass[preprint2]{aastex}

\usepackage{graphics}

\newcommand{\ngc}{NGC\,2359}
\newcommand{\hd}{HD\,56925}
\newcommand{\low}{$1 \rightarrow 0$}
\newcommand{\hig}{$2 \rightarrow 1$}
\newcommand{\kms}{km s$^{-1}$}

\newcommand{\hii}{{\sc Hii}}
\newcommand{\htwo}{H$_2$}

\newcommand{\amm}{NH$_3$}

\newcommand{\vlsr}{V$_{\rm LSR}$}

\shorttitle{Ammonia in NGC\,2359}
\begin{document}


\title{Shocked ammonia in the WR nebula NGC 2359}


\author{J.~R.~Rizzo\altaffilmark{1}, J.~Mart\'{\i}n-Pintado}
\affil{Observatorio Astron\'omico Nacional, Aptdo. Correos 1143, 28800
Alcal\'a de Henares, Spain} \email{jrizzo@oan.es, martin@oan.es}
\and
\author{C.~Henkel}
\affil{Max-Planck-Institut f\"ur Radioastronomie, Auf dem H\"ugel 69, 53121
       Bonn, Germany} \email{p220hen@mpifr-bonn.mpg.de}


\altaffiltext{1}{On leave of absence from Instituto Argentino de
Radioastronom\'{\i}a, Argentina}

%% ============  A B S T R A C T ======================================

\begin{abstract}

We report the detection of the (1,1) and (2,2) metastable lines of ammonia
(NH$_3$) in the molecular cloud associated with the Wolf--Rayet (WR) nebula
NGC\,2359. Besides the CO and H$_2$, this is the first molecule detected in 
the environs of a WR star. Width ($\Delta V_{\rm 1/2}$ = 3\,km\,s$^{-1}$) and
radial velocity ($V_{\rm LSR}\sim$ 54\,km\,s$^{-1}$) indicate that the
NH$_3$ lines arises from the  molecular cloud which is interacting with the WR
star. The rotational  temperature derived from the (1,1) and (2,2) line
intensity ratios is about 30\,K, significantly larger than the typical
kinetic temperature of the ambient gas of $\sim$  10\,K. The derived NH$_3$
abundance is $\sim 10^{-8}$. Linewidth, abundance and kinetic temperature can
be  explained if NH$_3$ is released from dust grain mantles to the gas phase by
 shocks produced by the expansion of the bubble created by the WR stellar
wind. We briefly discuss the implications of the detection of warm
NH$_3$ associated with a WR star in connection to the hot NH$_3$ emission
detected in the Galactic Center and in the nuclei of external galaxies.

\end{abstract}


%% =======  I N T R O D U C T I O N ==================================

\section{Introduction}

The interstellar medium surrounding Wolf--Rayet (WR) stars is expected to be
strongly disturbed by the fast stellar evolution with rapidly varying mass
loss rates, luminosities, and elemental composition of the ejecta. The
Wolf-Rayet (WR) star HD\,56925 --WR 7 in the catalogue of van der Hucht
(2001)-- excites the wind-blown bubble \ngc\ (Chu, Treffers \& Kwitter 1983).
The chemical enrichement (Esteban et al.\ 1990), the nearly spherical
morphology and the kinematics of \ngc\ suggest a  relatively recent origin for
this nebula, probably produced in the WR stage of the exciting star. The \hii\
region which surrounds the wind--blown bubble seems to be confined by
molecular material in the southern direction (Schneps et al.\ 1981). The
spatial distribution of the 1--0 (1) line of \htwo\ (St-Louis et al.\ 1998)
suggests that the \htwo\ vibrationally excited emission is produced by the
interaction of the WR star with the molecular material.

Recently Rizzo, Mart\'{\i}n-Pintado \& Mangum (2001, paper I) have mapped the
CO J = \low\ and J = \hig, and the $^{13}$CO J = \low\ emission over the whole
nebula. They found that the two velocity components toward the southern part
of the nebula have LSR radial velocities (\vlsr) of 54 and 67 \kms. The 67
\kms\ component has a narrow linewidth (1 \kms) indicating that this component
traces quiescent material. On the other hand, the component with \vlsr\ of 54
\kms\ is broad (5 \kms), surrounds the southern and eastern parts of \ngc\
(see Fig.~1a), and is spatially anticorrelated with the narrow component. The
broad linewidths and the spatial distribution of the 54 \kms\ component
indicate that it is affected by the interaction of the WR star with the
ambient material at 67 \kms. The shocks produced by the expansion of the
stellar wind bubble are likely the origin of the line broadening.

Molecular chemistry can be an important tool to firmly establish if the
molecular material in the broad component is  affected by the strong shocks
produced by the stellar wind from \hd. In particular, the ammonia (\amm)
fractional abundance changes from $10^{-10}$ in photodissociation regions to
$10^{-6}$ in shocked regions. This is due to the fact that the \amm\ is a very
fragile molecule, easily photodissociated by UV radiation (van Dishoeck 1988;
Roberge et al.\ 1991), but it can also be released from icy grain mantles by
shocks with moderate velocities (Flower \& Pineau-Des-Forets 1994). In this
letter we report the first detection of ammonia emission towards a WR nebula.
The rather high abundance of \amm\ and the kinetic temperature found in the
broad velocity component is in agreement with the idea that the molecular gas
has been shocked by the stellar wind from the WR star \hd.

%% ================= O B S E R V A T I O N S ==============================

\section{Observations and results}

The (1,1) and (2,2) metastable ammonia lines were observed using the
Effelsberg 100--m radiotelescope of the MPIfR in December 2000 and March 2001.
The half power beam width of the telescope at the rest frequency of the \amm\
lines, 23.7 GHz, was 40\arcsec. We used the new cooled dual channel HEMT
K--band receiver with a typical system temperature of 230\,K on a main beam
brightness temperature scale. An 8192--channel autocorrelator was used as the
backend. Both ammonia lines were observed simultaneously with a total
bandwidth of 20 MHz and a velocity resolution of 0.06 \kms. In order to
improve the S/N ratio, the spectra were later smoothed to a resolution of 0.5
\kms. The observations were made in a dual beam switching mode (switching
frequency 1 Hz) with a beam throw of 121\arcsec\ in azimuth. Pointing was
regularly checked and it was found to be within 5\arcsec. The data were
calibrated using the continuum emission of NGC\,7027 as the principal
calibrator, assuming a flux density of 5.4 Jy at 23.9 GHz (Ott et al.\ 1994).
Orion-KL (Hermsen et al.\ 1988) was used as a line calibrator. Both
calibrations agree to within 10\% and we conclude that line intensities on a
main beam temperature scale are accurate to within $\pm$15\%.

Figure 1a shows the spatial distribution of the $^{12}$CO J = \hig\ emission of
the broad component at \vlsr\ = 53 \kms\ in \ngc, overlaid on an optical
image of the WR nebula. The filled square indicates the location of the
maximum CO emission, which is also the position where the \amm\ data were
taken. Figs.\ 1b to 1e show the line profiles of the J = \low\ lines of
$^{12}$CO and $^{13}$CO, and the (1,1) and (2,2) lines of \amm, respectively.
All the lines were measured with similar beam sizes of 40\arcsec--50\arcsec.
Both ammonia lines are weak but clearly detected at a 7--8 $\sigma$ levels.
The observational parameters of the lines have been derived from gaussian fits
to the profiles of the (1,1) and (2,2) lines, and the results are shown in
Table 1. The radial velocities and linewidths are in good agreement with those
of the broad CO component at 53--54 \kms, indicating that both CO and \amm\
emission arises from the same region.

%%--------  T A B L E   1 ------------------------
\begin{table}[h]
\caption[]{Observational and physical parameters}
\begin{tabular}{l@{\qquad}l}
\noalign{\smallskip}
\hline
\hline
\noalign{\smallskip}

Parameter & Value \\

\noalign{\smallskip}
\hline
\noalign{\smallskip}

RA, Dec (B1950)      & $07^{\rm h}16^{\rm m}19^{\rm s},
-13\degr11\arcmin26\arcsec$ \\ \\
NH$_3$ (1,1) fit:   & \\
Peak emperature     & $53\pm 6$ mK \\
LSR velocity        & $53.1\pm 0.6$ \kms \\
Width profile       & $2.6\pm 0.6$ \kms \\
\\
NH$_3$ (2,2) fit:   & \\
Peak emperature     & $46\pm 7$ mK \\
LSR velocity        & $53.6\pm 0.6$ \kms \\
Width profile       & $3.0\pm 0.7$ \kms \\
\\
N(CO)               & $\sim 7.0 \times 10^{16}$ cm$^{-2}$ \\
N(NH$_3$)           & $\sim 6.8 \times 10^{12}$ cm$^{-2}$ \\
X(NH$_3$)           & $\sim 1.0 \times 10^{-8}$ \\
T$_{\rm rot}$       & $\sim$ 30 K \\

\noalign{\smallskip}
\hline
\end{tabular}
\end{table}

\resizebox{!}{17.2cm}{\includegraphics{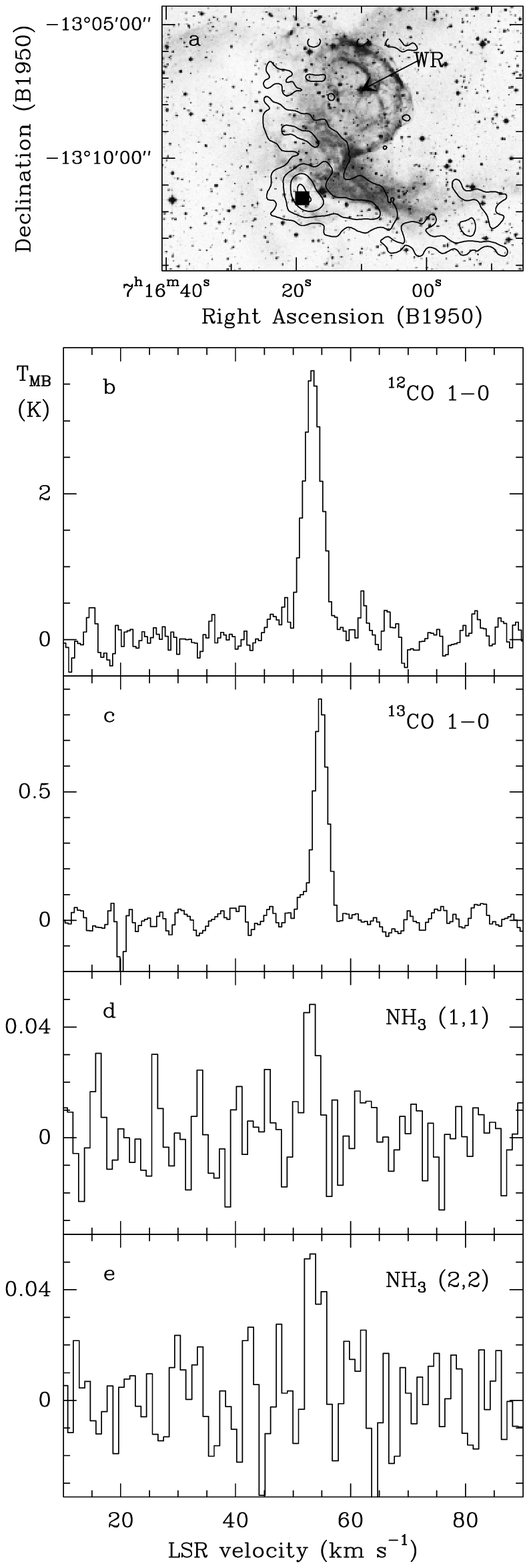}}

\figcaption[fig1.eps]{$^{12}$CO, $^{13}$CO and NH$_3$ emission toward the WR
nebula NGC 2359. The intensity distribution map in a) corresponds to the CO J =
\hig\ emission (see Paper I). The individual profiles are from the black
square plotted onto the clump to the southeast of the optical nebula. The
position of the WN4 star HD\,56925 is indicated by an arrow. In the profiles
from b) to e), the molecular transition is indicated at the top right corner.}

%%----------------------------------------------------------------

Assuming optically thin emission for both ammonia lines, we have derived the
(1,1) and (2,2) column densities given in Table 1. From the ratio of the (1,1)
and (2,2) column densities we estimate a rotational temperature of 30\,K.
Under typical conditions of the molecular gas, this rotational temperature
can be considered as a lower limit to the kinetic temperature of the gas
(Walmsley \& Ungerechts 1983). Using a kinetic temperature of 30\,K and
considering only the five lowest metastable (J = K) levels for the partition
function, we derive a lower limit to the total ammonia column density of $7
\times 10^{12}$ cm$^{-2}$. For a CO column density of $0.7 \times 10^{17}$
cm$^{-2}$ (paper I) and assuming a fractional CO abundance of $10^{-4}$, we
derive a \amm\ abundance of $1\pm 0.4 \times 10^{-8}$. This \amm\ abundance
is comparable to those measured in cold galactic molecular clouds (Benson
\& Myers 1983) and shocked regions (Mart\'{\i}n-Pintado \& Cernicharo
1987; Tafalla \& Bachiller 1995), but significantly higher than in PDR's
(Fuente et al.\ 1990).

%% ================= D I S C U S S I O N ===================================

\section{Discussion}

The broad widths and the radial velocities of the (1,1) and (2,2) \amm\ lines
indicate that the \amm\ emission arises from the gas which
is interacting with the WR nebula. Two different kinds of interaction between
an evolved massive star and the surrounding material are expected: UV
photodissociation and shocks produced by the stellar wind. It is well known
that the UV radiation from hot stars dissociates fragile molecules, such as
\amm, even up to visual extinctions of 5 mag (i.e., \htwo\ column densities of
$5 \times 10^{21}$ cm$^{-2}$). In this case, the photodissociation rate would
be of the order of $10^{-7}\ {\rm s}^{-1}$ (van Dishoeck 1988). Hence, the
relatively high \amm\ abundance is striking in view of the rather low \htwo\
column density of $\sim 10^{21}$ cm$^{-2}$ derived from the CO data. If the
total extinction between the star and the \amm\ region would be comparable to
the total extinction derived from the CO data, the \amm\ must be produced very
efficiently in order to compete with its photodissociation due to the WR star
radiation field. From the dust continuum emission measured by the IRAS HiRes
data at $60\mu$m and $100\mu$m (Aumann, Fowler \& Melnyk 1990; J.\ R.\ Rizzo
et al., in preparation), we have derived a dust temperature of 29\,K and a
column density in the range $2 - 5 \times 10^{16}$ cm$^{-2}$ towards the
position where \amm\ is detected. These values were obtained from the
color-corrected $60\mu$m and $100\mu$m emission maps smoothed to an angular
resolution of 2\arcmin, and assuming an emissivity proportional to $\lambda
^{-2}$. The dust grains were assumed to be spherical of size $0.1\mu$m and
with a density of 3 g\,cm$^{-3}$ (Hildebrand 1983). Assuming the standard
gas-to-dust ratio of 100, the total gas column density associated to the warm
dust at 29\,K is only $10^{18}$ cm$^{-2}$, at least two orders of magnitude
below the total gas column density derived from CO in paper I. This suggests
that basically all the dust associated with the molecular gas can not have
temperatures of $\ge$ 10\,K. The warm dust component measured by IRAS is
likely associated to the material affected by the UV radiation from the WR
star, shielding the warm gas from dissociation. The gas in the broad component
seems to be warmer than the dust.

The presence of shockfronts acting in this region could explain both the
\amm\ abundance and the gas kinetic temperature  higher than the dust
temperature in the broad component. A C--shock of 10--15 \kms\ would heat the
gas up to the temperatures derived from our data and release the \amm\ from the
dust grain icy mantles (Flower \& Pineau-Des-Forets 1994) to achieve the
observed abundance. Garc\'{\i}a-Segura \& Mac Low (1995a, 1995b) have modelled
the dynamical evolution of WR bubbles and predicted the presence of several
shockfronts running into the surrounding gas at the different stages of the
evolution of massive stars. When the WR stellar wind shocks the previous RSG
ejecta, an expansion velocity of $\sim$ 10--15 \kms\ is expected for the
shocked gas. This velocity is in good agreement with the velocity separation
of 12--13 \kms\ measured between the shocked and the quiescent gas (paper I).

The Galactic center region (GC) is known to contain large amounts of warm gas
with relatively high abundance of molecules like \amm, SiO and C$_2$H$_5$OH
extended over relatively large regions (H\"uttemeister et al.\ 1995;
Mart\'{\i}n-Pintado et al.\ 1997; Mart\'{\i}n-Pintado et al.\ 2001). This warm
gas seems to be associated with the cold dust, and a shock chemistry has been
claimed to explain the heating, the abundances and the densities of these
molecules in the GC. However, the origin of the widespread shocks is, so far,
unknown. Mart\'{\i}n-Pintado et al.\ (1999) have recently found that a large
fraction of the hot \amm\ in the Sgr B2 envelope is located in shells of 1--2
pc, expanding at 8--10 \kms. The large \amm\ abundance and the rather
high kinetic temperature found in the Sgr B2 envelope can be explained in
terms of a C-shock chemistry (Flower, Pineau Des Forets \& Walmsley 1995). 
Mart\'{\i}n-Pintado et al.\ (1999) proposed that these hot molecular shells
are the shocked layers produced by the strong winds of WR stars in the
Galactic Center. The detection of warm \amm\ associated with the expanding
shells produced by a WR stellar wind supports the possible origin for the warm
\amm\ in the Sgr B2 envelope. Although the densities in the GC clouds are
certainly higher than the density in \ngc, the ammonia abundances and the
expansion velocities in the GC shells are comparable to those found in \ngc.

Hot ammonia has also been observed in the nuclei of the external galaxies
IC\,342 (Martin \& Ho 1986; Ho et al.\ 1990) and Maffei 2 (Henkel et al.\
2000; Takano et al.\ 2000). The rotational temperatures of 50 K and 80 K for
IC\,342 and Maffei 2, respectively, are similar to those observed in the GC. In
both galaxy nuclei, the derived \amm\ abundances are also similar to the GC.
Furthermore, the gas temperatures derived from \amm\ are higher than the dust
temperature and the nucleus of IC\,342 has overall properties similar to those
of the GC region (Downes et al.\ 1992). In both galactic nuclei, IC\,342 and
Maffei 2, the gas heating is not due to recently formed stars but to large
scale shocks. As in the case of the GC, the origin of such shocks remains
unknown. Based on high angular resolution data (5\arcsec) from the nucleus of
IC\, 342, Ho et al.\ (1990) conclude that the morphology of the \amm\ emission
may be a distributed bump of small and hot molecular clouds surrounding young
OB stars. However, the \amm\ is hard to be heated by the OB field radiation
since it is easily photodissociated by the UV photons. It is worth noting that
this ammonia distribution would also be consistent with hot shells like those
observed by Mart\'{\i}n-Pintado et al.\ (1999) in the GC. Further studies of
the chemistry, physical conditions and kinematics of the molecular gas in the
environs of galactic evolved massive stars would provide unvaluable constraints
to the models for the impact of the evolution of massive stars onto the
interstellar medium.

\acknowledgments
This work has been partially supported by the Spanish CICYT grant
1FD1997-1442. JRR wish to thanks to the MPIfR staff for their kind help with
the observations at the 100m telescope. JRR is a fellow of the Conicet,
Argentina.

\end{document}